\documentclass[]{spie}  %>>> use for US letter paper
\usepackage{amsmath,amsfonts,amssymb}
\usepackage{graphicx}
\usepackage[colorlinks=true, allcolors=blue]{hyperref}
\usepackage{booktabs}
\usepackage{multirow}
\usepackage{color, colortbl}
\usepackage{xcolor}
\definecolor{Gray}{gray}{0.92}

\title{Per Clip and Per Bitrate Adaptation of the Lagrangian Multiplier in Video Coding }

\author{Daniel J. Ringis}
\author{Fran\c{c}ois Piti\'e}
\author{Anil Kokaram}
\affil{Sigmedia Group, Electronic and Electrical Engineering Dept., Trinity College Dublin, Ireland}

\authorinfo{For further information please contact D.J. Ringis :  ringisd@tcd.ie}

\begin{document} 
\maketitle

\begin{abstract}
In the past ten years there have been significant developments in optimization of transcoding parameters on a per-clip rather than per-genre basis. In our recent work we have presented per-clip optimization for the Lagrangian multiplier in Rate controlled compression, which yielded BD-Rate improvements of approximately 2\% across a corpus of videos using HEVC. However, in a video streaming application, the focus is on optimizing the rate/distortion tradeoff at a particular bitrate and not on average across a range of performance. 

We observed in previous work that a particular multiplier might give BD rate improvements over a certain range of bitrates, but not the entire range. Using different parameters across the range would improve gains overall. Therefore here we present a framework for choosing the best Lagrangian multiplier on a per-operating point basis across a range of bitrates. In effect, we are trying to find the para-optimal gain across bitrate and distortion for a single clip. In the experiments presented we employ direct optimization techniques to estimate this Lagrangian parameter path approximately 2,000 video clips. The clips are primarily from the YouTube-UGC dataset. We optimize both for bitrate savings as well as distortion metrics (PSNR, SSIM). 

%This idea is related to the Dynamic optimizer Framework proposed by Katsavounidis et al except here the dynamic nature of the optimization is across bitrate operating points instead of time. In addition, in this system we address a parameter (the Lagrangian multiplier) which is not normally explicitly related to content-specific tuning in this way.

\end{abstract}

% Include a list of keywords after the abstract 

\section{INTRODUCTION}
\label{sec:intro}  % \label{} allows reference to this section

With video content comprising over 80\% of all internet traffic~\cite{cisco}, there is an ever growing need for video compression for user generated content (UGC). The recent impact of the pandemic emphasises the importance of developing new techniques. The modern compression schemes H.265/6 (HEVC)~\cite{hevcOverview, zhang2019overview, bross2020versatile} and VP9 and AV1 \cite{mukherjee2013latest, chen2018overview}, have all been thrown into the spotlight.

Striking a balance between the rate and distortion of a video clip is a key challenge in video coding. Since 1998~\cite{sullivan1998rate}, it has been accepted practice to represent this trade-off in terms of a cost $J$ as follows:
\begin{equation}
    J = D + \lambda R \label{rd}
\end{equation}
Here, a distortion, $D$ and a rate $R$ are combined through the action of a Lagrangian multiplier, $\lambda$. The Lagrangian multiplier controls emphasis on either rate minimisation or quality improvement. This has an impact throughout the codec, as the idea is applied to many internal operations  e.g.  motion vectors, block type (Skipped/Intra/Inter), and bit allocation at the frame and clip levels~\cite{wiegand2001lagrange}.

The general idea which we have applied in our recent work \cite{pcsringis, SPIERingis, EIRingis} is to adjust $\lambda$ away from the codec default by using a constant $k$ as follows,
\begin{equation}
   \lambda_{\textrm{new}} = k \times \lambda_{\textrm{orig}}\label{kfactor}
\end{equation}
where $\lambda_{\textrm{orig}}$ is the default Lagrangian multiplier estimated in the video codec, and  $\lambda_{\textrm{new}}$ is the updated Lagrangian.

%Using this approach we had explored if a single value of $k$ was optimal across the corpus and found that improvements can be had in this adjustment. However, not every clip responded positively.

%Figure \ref{exampleBDk} shows BD-Rate vs $k$ for two different clips. There are two observations to be made. Firstly, the relationship between $k$ and BD-Rate varies substantially between clips. This reinforces the view that the traditional one size fits all approach is sub-optimal. Secondly, the general shape of the curves shows a global minimum. That minimum represents the maximum BD-Rate improvement available.

%\begin{figure}
%    \centering
%    \includegraphics[width=0.49\linewidth]{images/exp1.eps}
%    \includegraphics[width=0.49\linewidth]{images/exp3.eps}
%    \caption{BD-Rate vs $k = [0.2:0.2:3]$ for two different clips
%      (LiveMusic\_1080P-6d1a and NewsClip\_720P-7e56). The best performing BD-Rate gain is achieved at a
%      different value for $k$ in each clip. Hence per-clip optimization is
%      sensible. The curve shape shows a global minimum implying that classic
%      optimization strategies would be successful.}
%    \label{exampleBDk}
%\end{figure}

Using Equation \ref{kfactor}, we investigated in our most recent work\cite{PCSRingis} if a single value of $k$ would provide bitrate improvements across the corpus. We found that $k=0.786$ yields an average 0.63\% BD-rate improvement over the corpus, with 62\% of the clips showing an improvement. We then showed that a per clip optimization of $k$ can lead to an average 1.87\% BD-rate improvement, with 95\% of clips showing improvements.

In this paper, we expand on the Lagrangian Multiplier ($\lambda$) optimization idea by focusing on improving the bitrate savings at target bitrates in addition to each individual clip. Video clips are encoded at multiple bitrates using a variety of Lagrangian Multipliers in order to determine what is the best possible BD-Rate improvement for a clip at a given bitrate. Experiments on the same UGC for the HEVC codec, show that our per clip per operating point lambda optimization show that average of 2.67\% BD-Rate improvement, with 96\% showing improvements.
%\XXX 

%Contributions are: 1) estimate of the pareto ... 2) CRF, CBR and SSIM/PSNR, whereas it was only CRF/PNSR.
Contributions of this paper are 1) Estimation of the Pareto-Optimal BD-Rate improvement for HEVC and 2) Direct optimization of the Lagrangian Multiplier for constant bitrate encoding and SSIM as the distortion/quality metric where prior work focused solely on CRF encoding with PSNR.

\section{Background}

The rate distortion algorithm \cite{wiegand1996rate} establishes a balance between the quality of the media and the transmission or storage capacities of the medium. As mentioned above, seminal work of Sullivan and Wiegand \cite{sullivan1998rate} laid the foundation for an empirical approach to choosing an appropriate \(\lambda\) by notably establishing a relationship between the quantisation step size \(Q\) and the distortion \(D\) in a frame. Through minimising $J$ (Eq.\ref{rd}), this leads to  a relationship between \(\lambda\) and \(Q\) expressed as   $ \lambda = 0.85 \times Q^2 $. 
Updates to those experiments then yielded similar relationships for H.264 and H.265 (HEVC).  Because of the introduction of bi-directional (B) frames the constants are all different and three different relationships were established for each of the Intra (I), Predicted (P) and B frames as follows.
\begin{align}
   \lambda_{I} & =( 0.57 )2^{(Q-12)/3} \nonumber \\
\lambda_{P} & = (0.85) 2^{(Q-12)/3} \nonumber \\
\lambda_{B} & = ( 0.68 ) \max(2, \min(4, (Q - 12)/6))  2^{(Q-12)/3} \nonumber
\end{align}

\subsection{Per Clip Adaptive Lagrangian Multiplier}

There exists a limited amount of work on adaptation of $\lambda$ in the rate distortion equation. Most of this work focuses on creating prediction models based on other features of the video. These models are either machine learning systems based on video features extracted from the video content or either based on the Lambda-Q relationship which exists in the codec.

%can be focused on features gained from the video content or they can be based on the Lambda-Q relationship which exists in the codec.

%\XXX RE-READ THIS ENTIRE SECTION AND MAKE SURE I UNDERSTAND WHAT WAS IN THOSE PAPERS FROM MY WORDS! IF I CANT, I NEED TO RE-WRITE. give a little bit of context around each paragraph

\textbf{Machine Learning Approaches.}
Ma et al~\cite{ma2016adaptive} propose to use Support Vector Machine to determine \(k\). Their features include scalars representing spatial and temporal video information, as well as a texture feature based on a Gray Level Concurrence Matrix. Their focus was on {\em Dynamic textures} and they used the DynTex dataset of 37 sequences. They reported up to 2dB improvement in PSNR and 0.05 improvement in SSIM at equal bitrates. Hamza et al~\cite{hamza2019parameter} also take a classification approach but using gross scene classification into indoor/outdoor/urban/non-urban classes. They then used the same $k$ for each class. Their work used the Derfs dataset and reported up to 6\% BD-Rate improvement.Recently, John et al\cite{john2020rate} proposed a machine learning method to classify Rate Distortion characteristics of a clip in order to select the correct operating point for that clip. In this work, they cluster videos based on their RD-Curve operating points and determines the appropriate encoding parameters for each video based on the model developed.

\textbf{Predicting Lambda from QP.}
Zhang and Bull~\cite{zhang_bull} used a single feature ${D}_{P}/{D}_{B}$, the ratio between the MSE of P and B frames. This feature gives some idea of temporal complexity. Experiments based on the DynTex database yielded up to 7\% improvement in BD-Rate. They modified $\lambda$ implicitly by adjusting the quantiser parameter $Q$. Papadopoulos et al~\cite{Papadopoulos} exploited this and applied an offset to Q, in HEVC, based on the ratio of the distortion in the P and B frames. Each QP was updated from the previous Group of Pictures (GOP) using $ \textrm{QP} = a \times ({D}_{P}/{D}_{B}) - b $ where $a,b$ are constants determined experimentally.  This lead to an average BD-Rate improvement of 1.07\% on the DynTex dataset, with up to 3\% BD-Rate improvement achieved for a single sequence.
Yang et al~\cite{yang2017perceptual} used a combination of features instead of just the MSE ratio above. In their work, they used a perceptual content measurement $S$ to model $k$ with a straight line fit $ k = aS - b$. Here again $a,b$ were determined experimentally using a corpus of the Derfs dataset. The loss in complexity of the fit is compensated for by the increase in complexity of the feature. They report a BD-Rate improvement of up to 6.2\%.
 
%\XXX NETFLIX GOES HERE??

\subsection{Per Clip Direct optimization of the Lagrangian Multiplier on the YouTube-UGC Dataset\label{dopt}}

In recently published works \cite{EIRingis, SPIERingis, PCSRingis} we introduced the use of direct optimization to maximise BD-R and evaluated it on an expanded dataset, primarily based on the  YouTube-UGC dataset \cite{wang2019youtube}.  Multiple DASH segments (clips) of 5 seconds (150 frames) were created from each sequence. The dataset comprises a total of 9,746 video clips at varying resolutions with a wide range of video content, representative of typical usage. We present a summary of that previous work to set the context for our new contributions.

%By taking the best performing Lagrangian Multiplier at each bitrate, we can generate an encoding system which shows a great BD-Rate improvement over the wide range of operating points.

We minimise BD-Rate with respect to $k$ using direct optimization. The  BD-Rate \cite{bdrate} is defined as follows:
\begin{equation}\label{bjontegaardEqn}
    B_r(k) = \int_{D_1}^{D_2} (R_1(D) - R_k(D)) dD,
\end{equation}
where the integral is evaluated over the quality range $D_1..D_2$. $R_1(D), R_k(D)$ are the RD curves corresponding to the default ($k=1$) and the evaluated multiplier $k$ respectively. Each RD operating point is generated at a constant bitrate within a range which matches typical streaming media use cases.  The overall optimization process is then as follows.

\begin{enumerate}
    \item Generate an RD curve using $\lambda_{orig}$ and $R_t = {}$.  We use 5 operating points. 
    \item Define our BD-Rate objective function as specified above in equation \ref{bjontegaardEqn}. We use the same polynomial-log fit for evaluating the integral as recommended in \cite{bdrate}. %PSNR is used for the Distortion criterion.
    \item Starting from $k=1.0$,  minimise  $B_r(k)$, BD-Rate, wrt $k$ using any typical optimization routine.
\end{enumerate}
Note that for every evaluation of $B_r(k)$ , five (5) encodes are required as well as the subsequent BD-Rate calculation itself.

\begin{figure}
    \centering
    \includegraphics[height=0.6 \columnwidth]{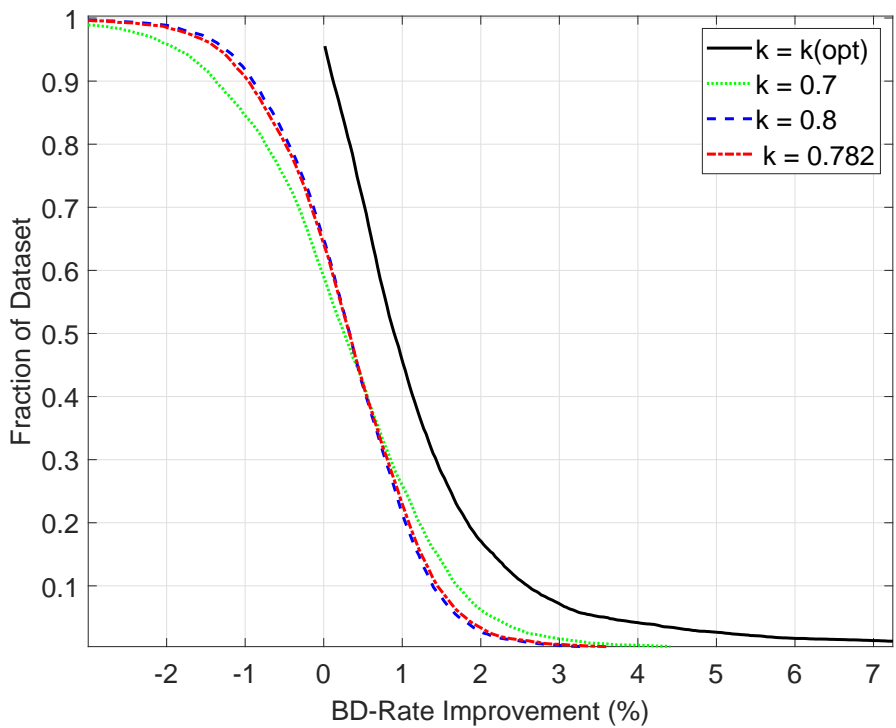}
    \caption{\label{SanityCheck}BD-Rate improvement vs fraction of dataset which achieves that improvement or better from \cite{PCSRingis}. We can see from this curve that the direct optimization method (k(opt)) was capable of achieving BD-Rate improvement in 95\% of the clips in the corpus. We can also read off of this graph what fraction of the clips had a 1\% or better BD-Rate improvement (47\%) and what fraction of the clips had a 5\% or better improvement (3\%). Also seen are the CDF plots when using $k$ = 0.7, 0.8 or 0.782. Ideally, we would want these plots to be as close as possible to the top right and coincide with the optimal per-clip plot (in black). We see for $k$ = 0.7, 0.8 or 0.782 positive BD-Rate improvements for 60-70\% of the clips but worse performance for 30-40\%.  }
\end{figure}

Figure \ref{SanityCheck} shows the BD-Rate improvement reported in \cite{PCSRingis} using direct optimization of the Lagrangian Multiplier. In this graph, we see the BD-Rate improvement on the x-axis and the fraction of the dataset which was able to achieve that BD-Rate improvement or better. Ideally, we would want this curve to be as close to the top right as possible, as that would indicate all clips achieved high BD-Rate improvements. The key result of this work is that 95\% of the clips had some BD-Rate improvement over the default Lagrangian Multiplier, with 46\% of them showing a BD-Rate improvement of 1\% or better. 

In addition to the direct optimization method, we proposed to investigate if $k=1$, as set in the MPEG standard, is indeed optimal for our corpus. We had found in \cite{PCSRingis} that for our corpus there is an improvement where $k$ is approximately 0.7:0.8, with a best average improvement coming at $k=0.782$.
We can see in Figure \ref{SanityCheck} that for $k=0.782$, about two-thirds of the clips have an improvement but it is worse for one-third of them. Overall, we get an average gain of 0.32\% across our corpus. This indicates that the Lagrangian Multiplier for HEVC may not be the best. This would imply that for HEVC, we should adjust the Lagrangian Multiplier to be $0.782 \times$ its current value. However, this may be specific to our corpus and the operating points which we are using. This gives further incentives for the need for the per clip approach.

%\begin{figure}
%    \centering
%    \includegraphics[height=0.6 \columnwidth]{images/directOpt.eps}
%    \caption{BD-Rate improvement vs fraction of dataset which achieves that improvement or better. We can see from this curve that the direct optimization method was capable of achieving BD-Rate improvement in 95\% of the clips in the corpus. We can also read off of this graph what fraction of the clips had a 1\% or better BD-Rate improvement (47\%) and what fraction of the clips had a 5\% or better improvement (3\%). }
%    \label{DirectOpt}
%\end{figure}

\section{Per Clip, Per Bitrate, Pareto optimization}

We believe that our past approach of optimizing across an entire range of an RD-Curve is beneficial, but sub-optimal. The Lagrangian multiplier has a different impact at different operating points, and there may be additional gains to be found by directly optimizing the Lagrangian Multiplier in smaller ranges of the RD-Curve. This will come at the cost of increased number of video encodes, but a clearer picture of what is the upper bound of bitrate savings using this per clip approach.

\begin{figure}
    \centering
    \includegraphics[width=0.8\columnwidth]{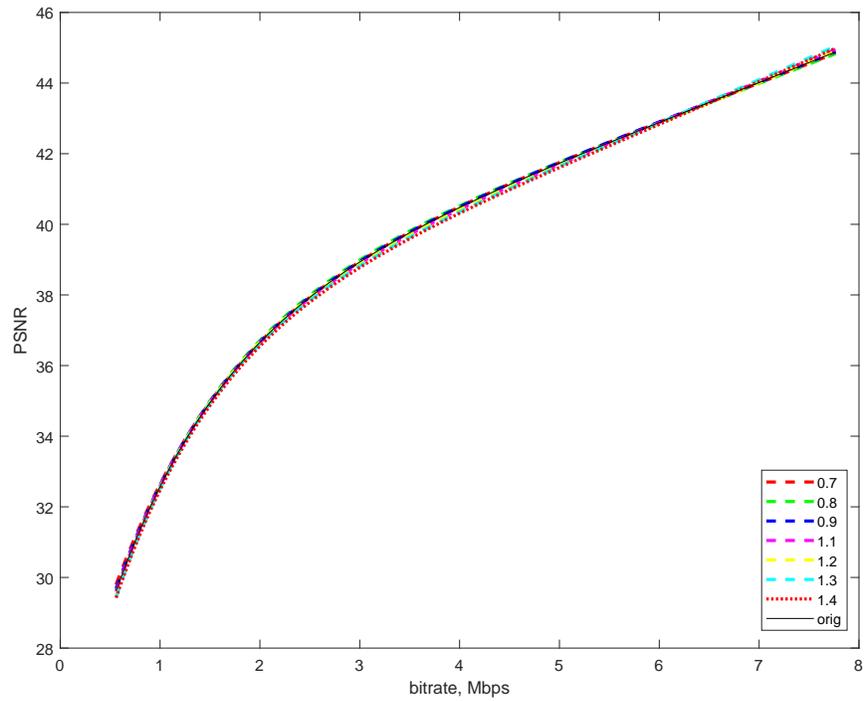}\\
    \includegraphics[width=0.8\columnwidth]{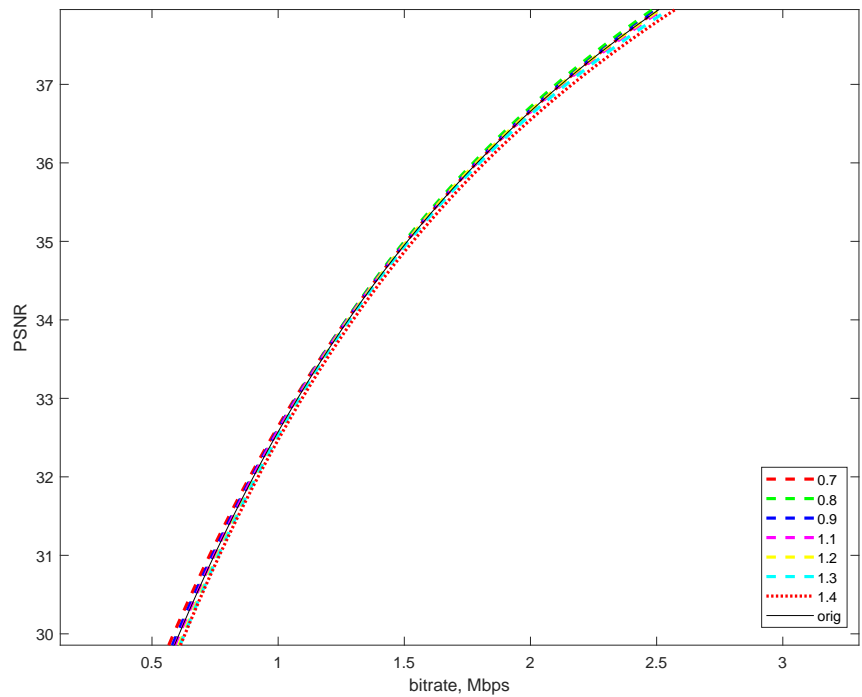}
    \caption{\label{paretoExample}RD-Curve for a single sequence (Top- full curve, Bottom - zoomed view). Each curve is a different value of $k$. We see in the range of 0.5-1Mbps $k=0.7$ is the best performer, whereas in the range > 2Mbps $k=0.8$ is the best performer. This trend is seen across the RD-Curve where at different intervals the value of $k$ which provides the best PSNR at a given bitrate changes. We take the supremum of each curve for a wide range of $k$}
\end{figure}

This target bitrate approach is suited to real world applications, as video content delivery systems typically provide videos at a few target bitrates as opposed to a full bitrate range. Our previous approach targets entire range but in practice a platform would be targeting a particular bitrate. So, gains that we reported are not optimal for each individual point. In Figure \ref{paretoExample} we show an example for a particular clip. We can see that in the low bit rate range 0.5-1.5Mbps, k=0.7 yields best quality but that k=0.8 yields best results for range 1.6-2.5Mbps. What we are looking for is the Pareto optimal of the set of all lambda values used when encoding a given clip at a particular bitrate. That is the curve that is made of the supremum of all these curves. 

To obtain this curve, we would need to find the optimal value of $k$ for a particular bitrate. However, this is a constrained optimization problem, a classic challenge for encoders that occurs often, and certainly hard to estimate perfectly. So instead, we take an empirical approach i.e. we generate a large amount of possible pairs $(bitrate, k)$ and sample the pareto surface directly from those points. As exhaustively encoding video at each bitrate with a large range of $k$ is not feasible, we need to establish a reasonable process to extract our pairs from our measurements in order to get our pareto-optimal curve.

In theory, our aim is this:
\begin{equation}
D_{\mathrm{Pareto}}(R) = \mathrm{sup}_k(D(R, k))
\end{equation}
however, in practice only a sparse sampling  of $\{D(R_i, k_j)\}_{i,j}$ is available from our experiments. 

Using the same interpolation scheme as used in the construction of the RD-curve \cite{bdrate}, we can estimate $\tilde{D}(R, k_j)$ for any value $\mathrm{bitrate}$. This solves our sparse sampling problem.  We have some confidence in the approximation $\tilde{D}(R, k_j)$ because the shape of the curve is well established to be approximately exponential\cite{bdrate}.

So our estimation becomes 
\begin{equation}
D_{\mathrm{Pareto}}(R) =  \mathrm{max}_j(\tilde{D}(R, k_j))
\end{equation}
% Figure YY, shows that this optimum can be narrow. 
However, taking the maximum for $k_j$ requires us to sample values for $k$ that are indeed close to the optimum. 
Interestingly, the direct optimization that we proposed in \cite{EIRingis} allows to focus the sampling of $k_i$ in the area of interest. However the previous work \cite{EIRingis} focuses on optimizing the BD rate improvement across the full range of bitrates for a given video clip.
In this work, we need to sample $k$ in such a way that it optimizes BD-Rate w.r.t. bitrate. That would in detail lead to many invocations of encoding, so 
we propose here to  define three bitrate ranges as detailed in the Table \ref{OperatingPoints}. The expectation is that the optimal value of $k$ for any given bitrate will then fall close enough to one of the three sets of values generated by our three experiments. In detail therefore we expand the range of candidates ($R$,$k_i$) by running 3 direct optimizations at low, medium and high bitrate ranges as specified in that table.

%As the operating points of these curves are discrete, we need to determine what would be the expected quality/distortion at a given bitrate for a given $k$ for each curve. By interpolating, we ensure that we do not have to do an exhaustive number of video encodes. We use exponential interpolation to determine the expected distortion at bitrates in between operating points. For each bitrate we take the point with the best quality as our best performer and combine them as seen in the example in Figure \ref{ExamplePareto}. We believe this to be our theoretical upper bound of bitrate improvement for each clip. The implementation of this and reported results are next.

\section{Experiments}

\subsection{YouTube-UGC Dataset}
Related work used a small corpus size (approximately 40 clips\cite{ma2016adaptive}\cite{zhang_bull} (up to 300 frames per clip)). Also the types of content used in previous corpora are not necessarily a good representation of modern material. For this work we use on HD Clips(720p and 1080p) clips from the recently published YouTube dataset \cite{wang2019youtube} representing 12 classes of video as specified by the YouTube team. Multiple DASH segments (clips) of 5 seconds (150 frames) were created from each sequence. This is a subset of the corpus used in \cite{SPIERingis, PCSRingis}.

\subsection{R-D Curve Operating Points}

%\XXX somewhere explain that, contrary to previous paper, we also measure PSNR/SSIM/CRF/CBR... Done in contributions.
The codebase for x265 was modified to take \(k\) as an argument. This allows the default $\lambda_{orig}$ to be modified according to equation \ref{kfactor}. We use the VideoLAN \cite{videoLAN} implementation of H.265, {\tt x265}\footnote{Version: 3.0+28-gbc05b8a91}. The following command invocations were used:

\textbf{CRF:} \texttt{x265 --input SEQ.y4m --crf <XX> --tune-<YYYY> --<YYYY> --csv-log-level 2 --csv DATA.csv --output OUT.mp4 }

\textbf{CBR:} \texttt{x265 --input SEQ.y4m --bitrate <ZZZZ> --tune-<YYYY> --<YYYY> --csv-log-level 2 --csv DATA.csv --output OUT.mp4 }

where {\tt SEQ}, {\tt DATA},  and  {\tt OUT}  are the filenames for the raw input file and output encoded video. We use the following operating points seen in Table \ref{OperatingPoints} for crf XX and target bitrate ZZZZ. YYYY is the distortion metric used, either PSNR or SSIM.

\begin{table}[h]
\caption{Operating points selected for the LOW, MED and HIGH ranges \label{OperatingPoints}}
\centering
\begin{tabular}{|l|l|l|}
\hline
              & \textbf{CRF} & \textbf{CBR}           \\ \hline
\textbf{LOW}  & 22:2:32      & 256kbps, 512kbps, 1Mbps, 2Mbps, 4Mbps \\ \hline
\textbf{MED}  & 27:2:37      & 1Mbps, 2Mbps, 4Mbps, 6Mbps, 8Mbps     \\ \hline
\textbf{HIGH} & 32:2:42      & 4Mbps, 6Mbps, 8Mbps, 10Mbps, 12Mbps   \\ \hline
\end{tabular}
\end{table}

We use Brent's method to directly optimize each range for a given clip. This takes approximately twelve (12) RD-Curve generations for each bitrate range. This leads to 180 video encodes for a given clip. Using the RD-Curves generated in each step of the direct optimization process, we fit an exponential curve with a step size of 1kbps. Using these curves we take the encode which has the maximum quality at each bitrate to form our Pareto-Optimal curve. The BD-Rate of the Pareto-Optimal curve, compared to the unmodified codec is calculated for each clip. Results across the ~2000 clip corpus are reported in the following section. Both SSIM and PSNR as distortion/quality metrics are are used.

% XXX METHOD
% Here we try to 

% \XXX rewrite As most applications focus on delivering a target bitrate to a user, we believe that this approach is better suited towards real world application.XXX

% XXX gist of it is: the sampling of the CRF/CBR values is important. 

%\XXX PARETO EXPLAINATION, INTERPOLATION OF THE RD CURVES

\section{Results}

\begin{figure}
    \centering
    \includegraphics[width=0.8\columnwidth]{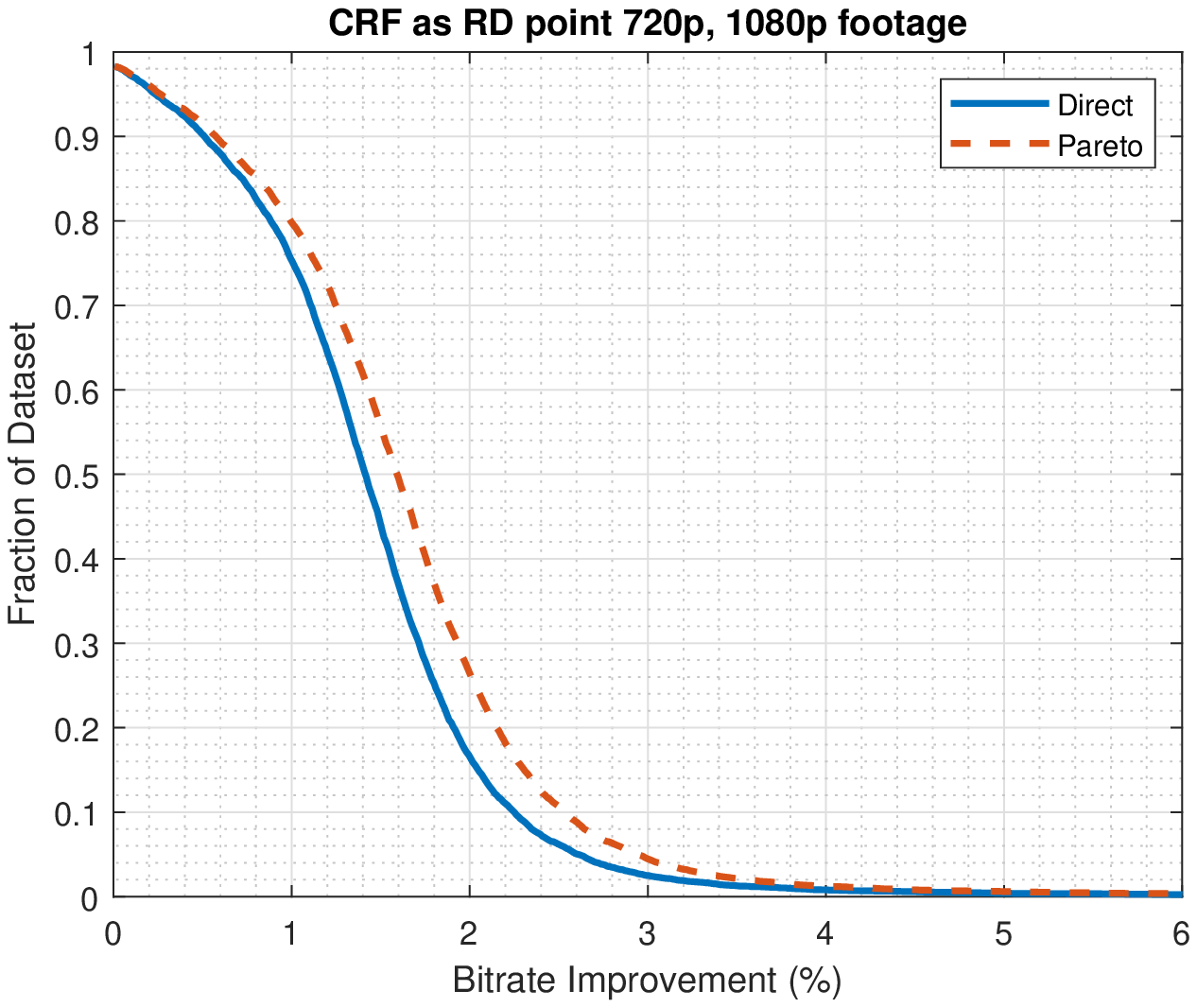}
    \includegraphics[width=0.8\columnwidth]{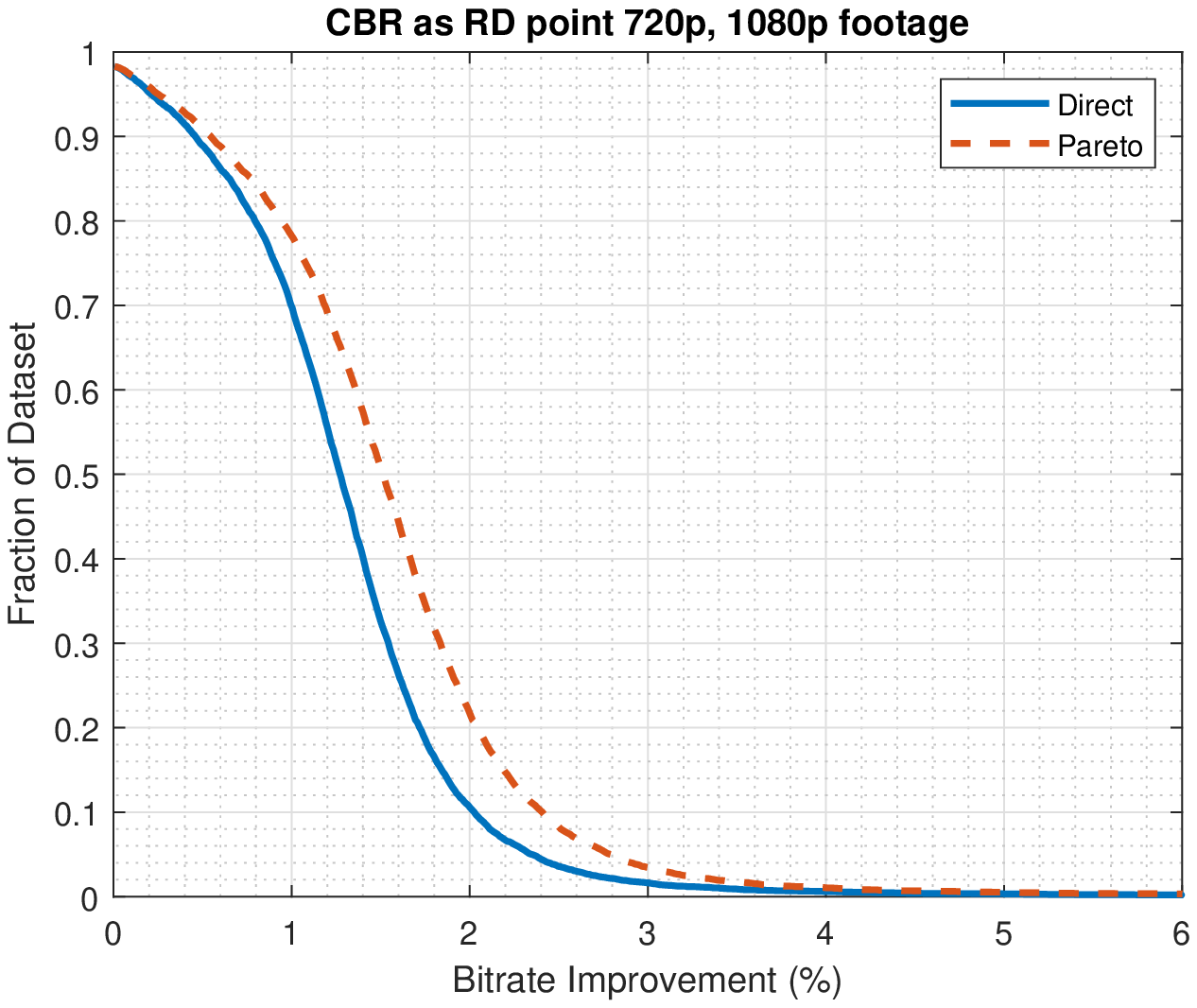}\\
    \caption{\label{fig:cdfspsnr} CDF plots of the BD-Rate improvement across our corpus using PSNR as the distortion metric. Top CRF operating point, Bottom CBR operating points. We can use these graphs to get an estimate of how much of the corpus has had BD-Rate improvements using the Direct optimizer system presented in \cite{EIRingis} (Direct) as well as using the Pareto-Optimal Direct optimization presented in this paper(Pareto)}
    \end{figure}

\begin{figure}
    \centering
    \includegraphics[width=0.8\columnwidth]{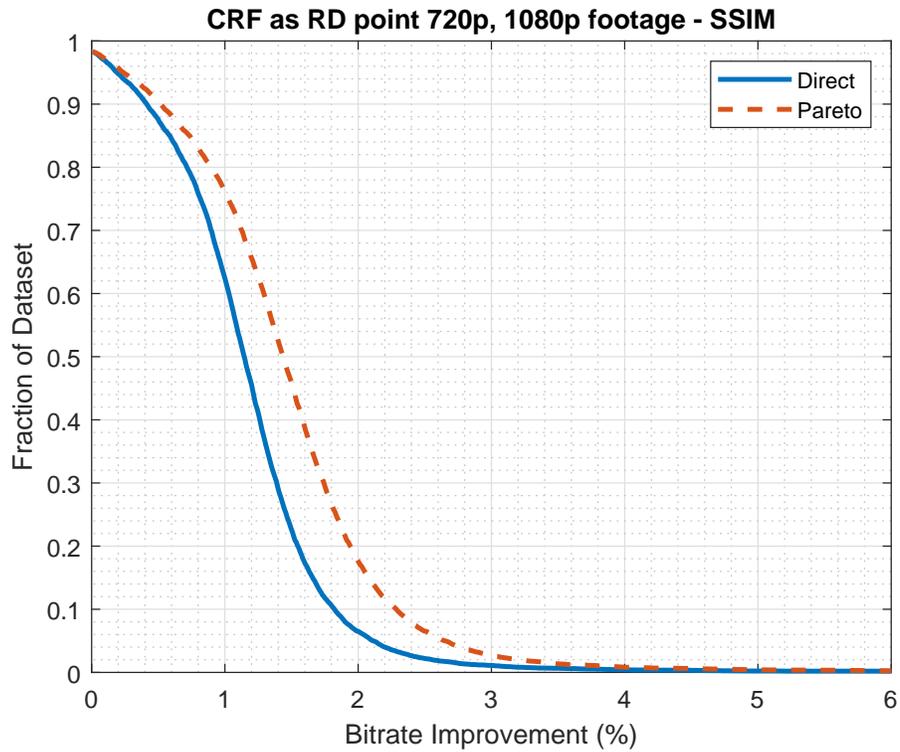}
    \includegraphics[width=0.8\columnwidth]{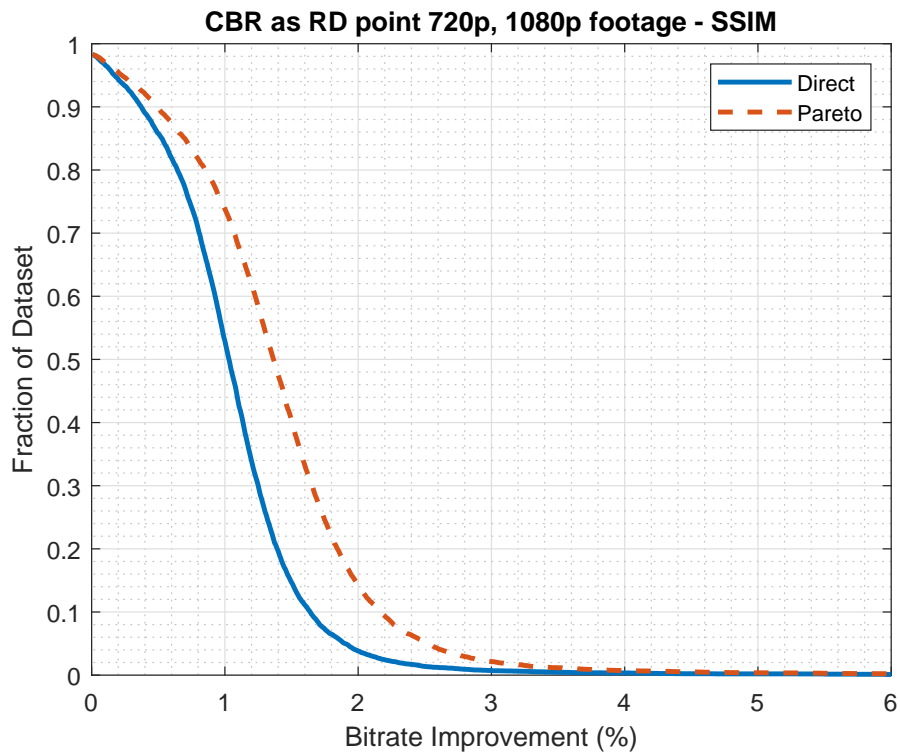}
    \caption{\label{fig:cdfsssim}CDF plots of the BD-Rate improvement across our corpus using PSNR as the distortion metric. Top CRF operating point, Bottom CBR operating points. We can use these graphs to get an estimate of how much of the corpus has had BD-Rate improvements using the Direct optimizer system presented in \cite{EIRingis} (Direct) as well as using the Pareto-Optimal Direct optimization presented in this paper(Pareto)}
\end{figure}

Each graph in Figures \ref{fig:cdfsssim} and \ref{fig:cdfspsnr} shows he cumulative distribution of the resulting maximum improvement (minimum BD-Rate using the optimal $k = k_D$). Direct \cite{EIRingis} refers to our past system where we directly optimize across a wide bitrate range and Pareto refers to the system presented in this paper. It is important to note that we are not comparing SSIM to PSNR nor CBR to CRF encoding, but investigating the improvement in gains by using the per bitrate system. Table \ref{GainsSummary} presents the key takeaways from these graphs. 


Table \ref{GainsSummary} also reports the {\em average positive BD-Rate improvement} as our {\em Average Final BD-Rate Gain}.
This is the gain measured when our system is used as a post-process or {\em optimized re-run} of the encoder given an initial encode with default settings. The best BD-rate between the initial encode and the re-run defines the final BD-Rate Gain. This is the most likely use of systems of this kind, and can even be considered as a two-pass system. This allows us to take advantage of any improvements in BD-Rate from Lagrangian multiplier optimization and default to the first default encode (unmodified Lagrangian Multiplier) in the case of worse performance. 

\begin{table}[t]
\centering
\caption{\label{GainsSummary} Summary of {\bf BD-R Gains} results. This shows the $\%$ of clips which have any improvement and $\>1\%$ improvement, as well as the average positive BD Rate improvement  }
\setlength{\tabcolsep}{4pt}
\begin{tabular}{rrrrrrr}
\toprule
\textbf{Operating}& \textbf{Distortion}& \textbf{Estimation} & \multicolumn{2}{c}{\bf Clips with BD-R Gain of} &  \textbf{Avg Final}   \\
\textbf{Point}&\textbf{ Metric} & \textbf{Method} & \textbf{$\geq 0\%$} & \textbf{$>1\%$} &  \textbf{Gain}   \\
% {} & \textbf{$\geq 0\%$} &  \textbf{$>1\%$} & \textbf{Gain} \\ 
\midrule
%Encoding & Distortion & Method & Clips with >0\% &
CBR & PSNR & Direct  & 96\% &   68\%   &  1.61\% \\
CBR & PSNR & Pareto  & 96\% &   79\%   &  2.67\% \\
CRF & PSNR & Direct  & 96\% &   74\%   &  2.24\% \\
CRF & PSNR & Pareto  & 96\% &   79\%   &  3.03\% \\
CBR & SSIM & Direct  & 98\% &   49\%    &  1.56\% \\
CBR & SSIM & Pareto  & 98\% &   74\%   &  3.02\% \\ 
CRF & SSIM & Direct  & 98\% &   59\%   &  1.05\% \\ 
CRF & SSIM & Pareto  & 98\% &  75\%   &  2.34\% \\

\bottomrule
\end{tabular}
\end{table}

The Pareto-Optimal method from this work has approximately 1-1.5\% BD-Rate improvement compared to the direct optimization method used in the past. This does come at the cost of at least three times the number of video encodes. It appears that this method yields similar improvements regardless of the distortion metric(PSNR/SSIM) or operating point(CBR/CRF) used. It is possible to get further gains with more than three encoding ranges, but we believe that as we near an exhaustive solution for the optimal Lagrangian multiplier for a given clip, we would be beyond the scope of reasonable computational cost for a small bitrate improvement. 

%We investigated whether there are specific ranges of k which impact the three regions, in Figure \ref{KHist} we see .....

%\XXX
%K HISTOGRAM PLOT

\section{Conclusion}

Overall, we are able to further increase the performance of the HEVC codec for a given clip by focusing on specific bitrate ranges. 
%We achieved up to a 1.5\% increase in performance, however, it was at the trade-off of 180 video encodes as opposed to 60 video encodes in the previous systems.
We achieve a further 1.5\% increase in performance compared to our previously reported system. However this increase in performance was at the expense of 3$\times$ the number of video encodes.Further  investigation is required into links between $k$, bitrate range and improvement found for a given clip. %investigation into the relationship between our adjustment  also found that the optimal Lagrangian Multiplier for a given clip widely varies at different ranges.

It is possible to get further gains with more than three encoding ranges, but we believe that as we near an exhaustive solution for the optimal Lagrangian multiplier for a given clip, we would be beyond the scope of reasonable computational cost for a small bitrate improvement. Future work would entail investigating the "stopping point" for this work, where the bitrate savings found are not worth the environmental and computational cost required to achieve them.

%\appendix    %>>>> this command starts appendixes
\acknowledgments % equivalent to \section*{ACKNOWLEDGMENTS}       
 
This work was supported in part by YouTube, Google and the Ussher Research Studentship from Trinity College.

\bibliography{main} % bibliography data in report.bib

\begin{thebibliography}{10}

\bibitem{cisco}
Cass, S., ``The age of the zettabyte cisco: the future of internet traffic is
  video [dataflow],'' {\em IEEE Spectrum}~{\bf 51}(3),  68--68 (2014).

\bibitem{hevcOverview}
{Sullivan}, G.~J., {Ohm}, J., {Han}, W., and {Wiegand}, T., ``Overview of the
  high efficiency video coding {HEVC} standard,'' {\em IEEE Transactions on
  Circuits and Systems for Video Technology}~{\bf 22},  1649--1668 (Dec 2012).

\bibitem{zhang2019overview}
Zhang, T. and Mao, S., ``An overview of emerging video coding standards,'' {\em
  GetMobile: Mobile Computing and Communications}~{\bf 22}(4),  13--20 (2019).

\bibitem{bross2020versatile}
Bross, B., Chen, J., Liu, S., and Wang, Y.-K., ``Versatile video coding (draft
  10),'' {\em ITU-T and ISO/IEC JVET-S2001}  (2020).

\bibitem{mukherjee2013latest}
Mukherjee, D., Bankoski, J., Grange, A., Han, J., Koleszar, J., Wilkins, P.,
  Xu, Y., and Bultje, R., ``The latest open-source video codec {VP9}-an
  overview and preliminary results,'' in [{\em 2013 Picture Coding Symposium
  (PCS)}{\nolinebreak\hspace{0.1em}]},   390--393, IEEE (2013).

\bibitem{chen2018overview}
Chen, Y., Murherjee, D., Han, J., Grange, A., Xu, Y., Liu, Z., Parker, S.,
  Chen, C., Su, H., Joshi, U., et~al., ``An overview of core coding tools in
  the av1 video codec,'' in [{\em 2018 Picture Coding Symposium
  (PCS)}{\nolinebreak\hspace{0.1em}]},   41--45, IEEE (2018).

\bibitem{sullivan1998rate}
Sullivan, G.~J. and Wiegand, T., ``Rate-distortion optimization for video
  compression,'' {\em IEEE signal processing magazine}~{\bf 15}(6),  74--90
  (1998).

\bibitem{wiegand2001lagrange}
Wiegand, T. and Girod, B., ``Lagrange multiplier selection in hybrid video
  coder control,'' in [{\em Proceedings 2001 International Conference on Image
  Processing}{\nolinebreak\hspace{0.1em}]},   {\bf 3},  542--545, IEEE (2001).

\bibitem{pcsringis}
Ringis, D.~J., Piti{\'e}, F., and Kokaram, A., ``Near optimal per-clip
  lagrangian multiplier prediction in hevc,'' in [{\em 2021 Picture Coding
  Symposium (PCS)}{\nolinebreak\hspace{0.1em}]},  (2021).

\bibitem{SPIERingis}
Ringis, D.~J., Piti{\'e}, F., and Kokaram, A., ``Per-clip adaptive lagrangian
  multiplier optimisation with low-resolution proxies,'' in [{\em Applications
  of Digital Image Processing XLIII}{\nolinebreak\hspace{0.1em}]},   {\bf
  11510},  115100E, International Society for Optics and Photonics (2020).

\bibitem{EIRingis}
Ringis, D.~J., Pitie, F., and Kokaram, A., ``Per clip lagrangian multiplier
  optimisation for ({HEVC}),'' {\em Electronic Imaging}~{\bf 2020}(12) (2020).

\bibitem{wiegand1996rate}
Wiegand, T., Lightstone, M., Mukherjee, D., Campbell, T.~G., and Mitra, S.~K.,
  ``Rate-distortion optimized mode selection for very low bit rate video coding
  and the emerging {H. 263} standard,'' {\em IEEE Transactions on Circuits and
  Systems for Video Technology}~{\bf 6}(2),  182--190 (1996).

\bibitem{ma2016adaptive}
Ma, C., Naser, K., Ricordel, V., Le~Callet, P., and Qing, C., ``An adaptive
  lagrange multiplier determination method for dynamic texture in {HEVC},'' in
  [{\em 2016 IEEE International Conference on Consumer Electronics-China
  (ICCE-China)}{\nolinebreak\hspace{0.1em}]},   1--4, IEEE (2016).

\bibitem{hamza2019parameter}
Hamza, A.~M., Abdelazim, A., and Ait-Boudaoud, D., ``Parameter optimization in
  {H. 265} rate-distortion by single frame semantic scene analysis,'' {\em
  Electronic Imaging}~{\bf 2019}(11),  262--1 (2019).

\bibitem{john2020rate}
John, S., Gadde, A., and Adsumilli, B., ``Rate distortion optimization over
  large scale video corpus with machine learning,'' {\em arXiv preprint
  arXiv:2008.12408}  (2020).

\bibitem{zhang_bull}
{Zhang}, F. and {Bull}, D.~R., ``Rate-distortion optimization using adaptive
  lagrange multipliers,'' {\em IEEE Transactions on Circuits and Systems for
  Video Technology}~{\bf 29},  3121--3131 (Oct 2019).

\bibitem{Papadopoulos}
{Papadopoulos}, M.~A., {Zhang}, F., {Agrafiotis}, D., and {Bull}, D., ``An
  adaptive {QP} offset determination method for {HEVC},'' in [{\em 2016 IEEE
  International Conference on Image Processing
  (ICIP)}{\nolinebreak\hspace{0.1em}]},   4220--4224 (Sep. 2016).

\bibitem{yang2017perceptual}
Yang, A., Zeng, H., Chen, J., Zhu, J., and Cai, C., ``Perceptual feature guided
  rate distortion optimization for high efficiency video coding,'' {\em
  Multidimensional Systems and Signal Processing}~{\bf 28}(4),  1249--1266
  (2017).

\bibitem{wang2019youtube}
{Wang}, Y., {Inguva}, S., and {Adsumilli}, B., ``Youtube {UGC} dataset for
  video compression research,'' in [{\em 2019 IEEE 21st International Workshop
  on Multimedia Signal Processing (MMSP)}{\nolinebreak\hspace{0.1em}]},   1--5
  (Sep. 2019).

\bibitem{bdrate}
Bjontegaard, G., ``Calculation of average {PSNR} differences between rd curves;
  {VCEG-M33},'' tech. rep. (2001).

\bibitem{videoLAN}
VideoLAN, ``{x265, the free {H.265} encoder}.''
\newblock https://www.videolan.org/developers/x265.html.

\end{thebibliography}
\bibliographystyle{spiebib} % makes bibtex use spiebib.bst

\end{document}